\documentclass[twocolumn,showpacs,aps,floatfix]{revtex4}
\usepackage{amsmath,amssymb,eucal,graphicx}
\begin{document}
\title{How to Choose a Champion}
\author{E.~Ben-Naim$^1$ and N.~W.~Hengartner$^2$}
\affiliation{$^1$Theoretical Division and Center for Nonlinear Studies,
$^2$Computational, Computer, and Statistical Sciences Division,
Los Alamos National Laboratory, Los Alamos, New Mexico 87545 USA}
\begin{abstract}
  League competition is investigated using random processes and
  scaling techniques. In our model, a weak team can upset a strong
  team with a fixed probability.  Teams play an equal number of
  head-to-head matches and the team with the largest number of wins is
  declared to be the champion.  The total number of games needed for
  the best team to win the championship with high certainty, $T$,
  grows as the cube of the number of teams, $N$, i.e., $T\sim
  N^3$. This number can be substantially reduced using preliminary
  rounds where teams play a small number of games and subsequently,
  only the top teams advance to the next round. When there are $k$
  rounds, the total number of games needed for the best team to emerge
  as champion, $T_k$, scales as follows, $T_k\sim N^{\gamma_k}$ with
  $\gamma_k=[1-(2/3)^{k+1}]^{-1}$. For example,
  $\gamma_k=9/5,27/19,81/65$ for $k=1,2,3$.  These results suggest an
  algorithm for how to infer the best team using a schedule that is
  linear in $N$.  We conclude that league format is an ineffective 
  method of determining the best team, and that sequential elimination
  from the bottom up is fair and efficient.
\end{abstract}
\pacs{02.50.-r, 01.50.Rt, 05.40.-a, 89.75.Da}
\maketitle

\section{Introduction}

Competition is ubiquitous in physical, biological, sociological, and
economical processes.  Examples include ordering kinetics where large
domains grow at the expense of small ones \cite{gss,ajb}, evolution
where fitter species thrive at the expense of weaker species
\cite{sjg}, social stratification where humans vie for social status
\cite{btd,msk,bvr}, and the business world where companies compete for
market share \cite{ms,ra}.

The world of sports provides an ideal laboratory for modeling
competition because game data are accurate, abundant, and
accessible. Moreover, since sports competitions are typically
head-to-head, sports can be viewed as an interacting particle system,
enabling analogies with physical systems that evolve via binary
interactions \cite{tmp,pn,brv,ta}. For instance, sports nicely
demonstrate that the outcome of a single competition is not
predictable \cite{tl,bvr1}.  Over the past century the lower seeded
team had an astounding $44\%$ chance of defeating a higher seeded team
in baseball \cite{bvr1}. The same is true for other competitions in
arts, science, and politics. This inherent randomness has profound
consequences.  Even after a long series of competitions, the best team
does not always finish first.

To understand how randomness affects the outcome of multiple
competitions, we study an idealized system. In our model league, there
are $N$ teams ranked from best to worst, so that in each match there
is a well-defined favorite and underdog. We assume that the weaker
team can defeat the stronger team with a fixed probability.  Using
random walk properties and scaling techniques analogous to those used
in polymer physics \cite{dg,de}, we study the rank of the champion as
a function of the number of teams and the number of games. We find
that a huge number games, $T\sim N^3$, is needed to guarantee that the
best team becomes the champion.

We suggest that a more efficient strategy to decide champions is to
set up preliminary rounds where a small number of games is played and
based on the outcome of these games, only the top teams advance to the
next round. In the final championship round, $M$ teams play a
sufficient number of $M^3$ games to decide the champion. Using $k$
carefully constructed preliminary rounds, the required number of
games, $T_k$, can be reduced significantly
\begin{equation}
\label{tk}
T_k\sim N^{\gamma_k}\qquad \gamma_k=\frac{1}{1-(2/3)^{k+1}}.
\end{equation}
Remarkably, it is possible to approach the optimal limit of linear
scaling using a large number of preliminary rounds.

\section{League competition}

Our model league consists of $N$ teams that compete in head-to-head
matches. We assume that each team has an innate strength and that no
two teams are equal. The teams are ranked from $1$ (the best team) to
$N$ (the worst team). This ranking is fixed and does not evolve with
time.  The teams play a fixed number of head-to-head games, and each
game produces a winner and a loser. In our model, the stronger (lower
seed) team is considered to be the favorite and the weaker (higher
seed) team is considered to be the underdog. The outcome of each match
is stochastic: the underdog wins with the upset probability $0<q<1/2$
and the favorite wins with the complementary probability $p=1-q$. The
team with the largest number of wins is the champion.

Since the better team does not necessarily win a game, the best team
does not necessarily win the championship.  In this study, we address
the following questions: How many games are needed for the best team
to finish first?  What is the typical rank of a champion decided by a
relatively small number of games?  What is the optimal way to choose a
champion?

We answer these questions using scaling techniques. Consider the $n$th
ranked team with $1\leq n\leq N$.  This team is inferior to a fraction
$\frac{n-1}{N-1}$ of the $N-1$ remaining teams and superior to a
fraction $\frac{N-n}{N-1}$ of the teams. Therefore, the probability
$P_n$ that this team wins a game against a randomly chosen opponent is
a linear combination of the probabilities $p$ and $q$,
\begin{equation}
\label{pn1}
P_n=p\,\frac{N-n}{N-1}+q\,\frac{n-1}{N-1}.
\end{equation}
Using $p=1-q$, the probability $P_n$ can be rewritten as follows
\begin{equation}
\label{pn}
P_n=p-(2p-1)\frac{n-1}{N-1}.
\end{equation}
The latter varies linearly with rank: it is largest for the best team,
$P_1=p$, and smallest for the worst team, $P_N=q$.

Now, suppose that the $n$th team plays $t$ games, each against a
randomly chosen opponent. The number of wins it accumulates, $w_n(t)$,
is a random quantity that grows as follows 
\begin{equation}
w_n(t+1)=
\begin{cases}
w_n(t)+1&{\rm with\ probability\ }P_n\\
w_n(t)  &{\rm with\ probability\ }1-P_n.\\
\end{cases}
\end{equation}
The initial condition is $w_n(0)=0$. The number of wins performs a
biased random walk and as a result, when the number of games is large,
the quantity $w_n(t)$ is well-characterized by its average
\hbox{$W_n(t)=\langle w_n(t)\rangle$} and its standard deviation
$\sigma_n(t)$, defined via \hbox{$\sigma^2_n(t)=\langle
w_n^2(t)\rangle-\langle w_n(t)\rangle^2$}.  Here, the brackets denote
averaging over infinitely many realizations of the random
process. Since the outcome of a game is completely independent of all
other games, the average number of wins and the variance in the number
of wins are both proportional to the number of games played
\begin{subequations}
\begin{align}
\label{vn}
W_n(t)&=P_n\,t\\
\label{wn}
\sigma^2_n(t)&=P_n(1-P_n)\,t.
\end{align}
\end{subequations}
Both of these quantities follow from the behavior after one game:
since $w_n(1)=1$ with probability $P_n$ and $w_n(1)=0$ with
probability $1-P_n$, then \hbox{$\langle w_n(1)\rangle=\langle
w_n^2(1)\rangle=P_n$}.  Moreover, the distribution of the number of
wins is binomial and for large $t$, it approaches a Gaussian, fully
characterized by the average and the standard deviation \cite{nvk}.

The quantities $W_n$ and $\sigma_n$ can be used to understand key
features of this system.  Let us assume that each team plays $t$ games
against randomly selected opponents and compare the best team with the
$n$th ranked team. Since $P_1>P_n$, the best team accumulates wins at
a faster rate, and after playing sufficiently many games, the best
team should be ahead. However, since there is a diffusive-like
uncertainty in the number of wins, \hbox{$\sigma_n\sim \sqrt{t}$}, it
is possible that the $n$th ranked team has more wins when $t$ is
small.  The number of wins of the $n$th team is comparable with that
of the best team as long as $W_1(t)-W_n(t)\propto \sigma_1(t)$, or
\begin{equation}
\label{compare}
(2p-1)\,\frac{n-1}{N-1}\,t\propto \sqrt{t}.
\end{equation}
Since the diffusion coefficient $D_n=P_n(1-P_n)$ in (\ref{wn}) varies
only weakly with $n$, $pq\leq D_n\leq 1/4$, this dependence is tacitly
ignored. When these two teams have a comparable number of wins, they
have comparable chances to finish first. Hence, Eq.~(\ref{compare})
yields the characteristic rank of the champion, $n_*$, as a function
of the number of teams $N$ and the number of games $t$
\begin{equation}
\label{nt}
n_*\sim \frac{N}{\sqrt{t}}.
\end{equation}
Since we are primarily interested in the behavior as a function of $t$
and $N$, the dependence on the probability $p$ is henceforth left
implicit. As expected, the champion becomes stronger as the number of
games increases (recall that small $n$ represents a stronger team). By
substituting $n_*\sim 1$ into (\ref{nt}), we deduce that the total
number of games, $t_*$, needed for the best team to win is $t_*\sim
N^2$.

Since each of the $N$ teams plays $t_*\sim N^2$ games, the total
number of games required for the best team to emerge as the champion
with high certainty grows as the cubic power of the number of teams,
\begin{equation}
\label{tn}
T\sim N^3.
\end{equation}
This result has significant implications. In most sports leagues, two
teams face each other a fixed number of times, usually once or twice.
The corresponding total number of $\sim N^2$ games, is much smaller
than (\ref{tn}).  In this common league format, the typical rank of
the champion scales as $n_*\sim \sqrt{N}$.  Such a season is much too
short as it enables weak teams to win championships.  Indeed, it is
not uncommon for the top two teams to trade places until the very end
of the season or for two teams to tie for first, a clear indication
that the season length is too short.

We may also consider the probability distribution $Q_n(t)$ for the
$n$th ranked team to win after $t$ games. We expect that the scale
$n_*$ characterizes the entire distribution function,
\begin{equation}
\label{qn-scaling}
Q_n\sim \frac{1}{n_*}\,\psi\left(\frac{n}{n_*}\right).
\end{equation}
Assuming $\psi(0)$ is finite, the probability that the best team wins
scale as follows, $Q_1\sim 1/n_*$. This quantity first grows,
\hbox{$Q_1(t)\sim\sqrt{t}/N$} when $t\ll N^2$, and then, it saturates,
$Q_1(t)\approx 1$ when $t\gg N^2$.

The likelihood of major upsets is quantified by the tail of the
scaling function $\psi(z)$. Generally, the champion wins $pt$ games
(we neglect the diffusive correction). The probability that the
weakest team becomes champion by reaching that many wins is
\hbox{$Q_N(t)\sim {t\choose pt}q^{pt}p^{qt}\sim (q/p)^{(p-q)t}$} where
the asymptotic behavior follows from the Stirling formula
\hbox{$t!\sim t\ln t-t$}.  We conclude that the probability of the
weakest team winning decays exponentially with the number of games,
$Q_N(t)\sim \exp(-\,{\rm const}\times t)$. Yet, from (\ref{qn-scaling})
and (\ref{nt}), $Q_N(t)\sim \psi\left(\sqrt{t}\,\right)$, and therefore,
the tail of the probability distribution is Gaussian
\begin{equation}
\label{tail}
\psi(z)\sim \exp\left(-\,{\rm const}\times z^2\right)
\end{equation}
as $z\to\infty$ thereby implying that upset champions are extremely
improbable. We note that single-elimination tournaments produce upset
champions with a much higher probability because the corresponding
distribution function has an algebraic tail \cite{brv}.  We conclude
that leagues have a much narrower range of outcomes and in this sense,
leagues are more fair than tournaments.

\section{Preliminary Rounds}

With such a large number of games, the ordinary league format is
highly inefficient.  How can we devise a schedule that produces the
best team as the champion with the least number of games? The answer
involves preliminary rounds. In a preliminary round, teams play a
small number of games and only the top teams advance to the next
round.

Let us consider a two stage format. The first stage is a preliminary
round where teams play $t_1$ games and then, the teams are ranked
according to the outcome of these games.  The top $M\ll N$ teams
advance to the final round \cite{clean}, and the rest are
eliminated. The final championship round proceeds via a league format
with plenty of games to guarantee that the best team ends up at the
top .

We assume that the number of teams advancing to the second round grows
sub-linearly
\begin{equation}
\label{m-def}
M\sim N^{\alpha_1},
\end{equation}
with $\alpha_1<1$. Of course, we better not eliminate the best team.
The number of games $t_1$ required for the top team to finish no worse
than $M$th place is obtained by substituting $n_*\sim M$ into
(\ref{nt}), $t_1\sim N^2/M^2$.  Since each of the $N$ teams plays
$t_1$ games, the total number of games in the preliminary round is of
the order \hbox{$Nt_1\sim N^3/M^2\sim N^{3-2\alpha_1}$}. Directly from
(\ref{tn}), the number of games in the final round is $M^3\sim
N^{3\alpha_1}$. Adding these two contributions, the total number of
games, $T_1$, is
\begin{equation}
\label{t1-eq}
T_1\sim N^{3-2\alpha_1}+N^{3\alpha_1}.
\end{equation}
This quantity grows algebraically with the number of teams, $T_1\sim
N^{\gamma_1}$ with $\gamma_1={\rm max}(3-2\alpha_1,3\alpha_1)$ and
this exponent is minimal, $\gamma_1=9/5$, when
\begin{equation}
\label{alpha1}
\alpha_1=3/5.
\end{equation}
Consequently, $t_1\sim N^{4/5}$.

Thus, it is possible to significantly improve upon the ordinary league
format using a two-stage procedure.  The first stage is a preliminary
round in which each of the $N$ teams plays $t_1 \sim N^{4/5}$ games
and then the top $M\sim N^{3/5}$ teams advance to the final round. The
rest of the teams are eliminated.  The first preliminary round
requires $N^{9/5}$ games. In the final round the remaining teams play
in a league with each of the possible ${M\choose 2}$ pairs of teams
playing each other $M$ times. Again the number of games is $N^{9/5}$
so that in total,
\begin{equation}
T_1\sim N^{9/5}
\end{equation}
games are played. This is a substantial improvement over ordinary
$N^3$ league play.

Multiple preliminary rounds further reduce the number of games.
Introducing an additional round, there are now three stages: the first
preliminary round, the second preliminary round, and the championship
round. Out of the first round $N^{\alpha_2}$ teams proceed to the
second round and then, $N^{\alpha_1\alpha_2}$ teams proceed to the
championship round. The total number of games $T_2$ is a
straightforward generalization of (\ref{t1-eq})
\begin{equation}
\label{t2-eq}
T_2\sim N^{3-2\alpha_2}+N^{\alpha_2(3-2\alpha_1)}+N^{3\alpha_1\alpha_2}.
\end{equation}
These three terms account respectively for the first round, the second
round, and the final round. The first term is analogous to the first
term in (\ref{t1-eq}), and the last two terms are obtained by
replacing $N$ with $N^{\alpha_2}$ in (\ref{t1-eq}).  The total number
of games is minimal when all three terms are of the same
magnitude. Comparing the last two terms gives $3-2\alpha_1=3\alpha_1$
and therefore, (\ref{alpha1}) is recovered.  Comparing the first two
terms gives
\begin{equation}
\label{alpha2-eq}
3-2\alpha_2=\alpha_2(3-2\alpha_1).
\end{equation}
Thus, $\alpha_2=15/19$ and since $\alpha_2>\alpha_1$, the first
elimination is less drastic then the second one.  The total number of
games, $T_2\sim N^{27/19}$, represents a further improvement.

These results indicate that it is possible to systematically reduce
the total number of games via successive preliminary rounds that lead
to the final championship round. In the most general case, there are
$k$ preliminary rounds in addition to the final round. The number of
teams advancing to the second round, $M_k$, grows as follows
\begin{equation}
\label{mk}
M_k\sim N^{\alpha_k}.
\end{equation}
From (\ref{alpha2-eq}), the exponent $\alpha_k$ obeys the recursion
relation $3-2\alpha_{k+1}=\alpha_{k+1}(3-2\alpha_k)$ or equivalently,
\begin{equation}
\label{alpha-eq}
\alpha_{k+1}=\frac{3}{5-2\alpha_k}.
\end{equation}
By using $\alpha_1=3/5$ we deduce the initial element in this series,
$\alpha_0=0$.  Introducing the transformation $\alpha_k=a_k/a_{k+1}$
reduces (\ref{alpha-eq}) to the Fibonacci-like recursion
\hbox{$3a_{k+2}=5a_{k+1}-2a_k$}. The general solution of this equation
is \hbox{$a_k=A\,r_1^k+B\,r_2^k$} where $r_1=1$ and $r_2=2/3$ are the
two roots of the quadratic equation $3r^2=5r-2$. The coefficients
follow from the zeroth element: $\alpha_0=0$ implies $a_0=0$ and
consequently, $a_k=A\left[1-(2/3)^k\right]$. Therefore,
\begin{equation}
\label{alpha-sol}
\alpha_k=\frac{1-\left(2/3\right)^k}{1-\left(2/3\right)^{k+1}}.
\end{equation}
The exponent $\alpha_k\approx 1-\frac{1}{3}\left(\frac{2}{3}\right)^k$
(for $k \gg 1$) decreases exponentially to one (Table 1).  This means
that the number of teams advancing from the first to the second
preliminary round is increasing with the total number of preliminary
rounds played. Nonetheless, the fraction of teams that are eliminated
$1-N^{\alpha_k-1}$ converges to one as $N\to\infty$.  Hence, nearly
all of the teams are eliminated in large leagues.

\begin{table}[t]
\begin{tabular}{|c|c|c|c|c|c|c|c|}
\hline
\,$k$\,&\,0\,&\,1\,&\,2\,&\,3\,&\,4\,&\,5\,&$\,\infty$\,\\
\hline
$\alpha_k$&0&$\frac{3}{5}$&$\frac{15}{19}$&$\frac{57}{65}$&$\frac{195}{211}$&$\frac{633}{665}$&1\\
$\beta_k$ &1&$\frac{4}{5}$&$\frac{8}{19} $&$\frac{16}{65}$&$\frac{32}{211} $&$\frac{64}{665} $&0\\
$\gamma_k$&3&$\frac{9}{5}$&$\frac{27}{19}$&$\frac{81}{65}$&$\frac{243}{211}$&$\frac{729}{665}$&1\\
\hline
\end{tabular}
\caption{The exponents $\alpha_k$, $\beta_k$, and $\gamma_k$
characterizing $M_k$, the number of teams advancing from the first
round, $t_k$, the number of games played by a team in the first round,
and $T_k$, the total number of games, as a function of the number of
preliminary rounds $k$.}
\end{table}

The number of games played by a team in the first round, $t_k$,
follows from (\ref{mk})
\begin{equation}
\label{tauk}
t_k\sim N^{\beta_k},\qquad \beta_k=2(1-\gamma_k).
\end{equation}
Since $\beta_k\to 0$ as \hbox{$k\to\infty$}, only a small number of
games is played in the opening round. Using $T_k\sim Nt_k$, we arrive
at our main result (\ref{tk}) where $\gamma_k=3-2\alpha_k$.
Surprisingly, the total number of games is roughly linear in the
number of teams
\begin{equation}
\label{linear}
T_k\sim N,
\end{equation}
when a large number of preliminary rounds is used, i.e.,
\hbox{$k\to\infty$} \cite{small}.  Clearly, this linear scaling is
optimal since every team must play at least once. The asymptotic
behavior $\gamma_k\approx 1+\left(\frac{2}{3}\right)^{k+1}$ implies
that in practice, a small number of preliminary round suffices.  For
example, $\gamma_4=\frac{243}{211}=1.15165$ (Table I).

We emphasize that in a $k$-round format, the top $N^{\alpha_k}$ teams
proceed to the second round, out of which the top
$N^{\alpha_{k-1}\alpha_k}$ teams proceed to the third round, and so
on. The number of teams proceeding from the $k$th round to the
championship round is $M\sim N^{\alpha_1\alpha_2\cdots\alpha_k}$. From
(\ref{linear}) and $T\sim M^3$, the size of the championship round
approaches
\begin{equation}
\label{M-scaling}
M\sim N^{1/3}
\end{equation}
as $k\to \infty$. This is the optimal size of a playoff that produces
the best champion using the least number of games.

\section{Numerical Simulations} 

Our scaling analysis is heuristic: we assumed that $N$ is very large
and we ignored numerical constants.  To verify the applicability of
our asymptotic results to moderately sized leagues, we performed
numerical simulations with $N$ teams that play an equal number of $t$
games against randomly selected opponents. The outcome of each game is
stochastic: with probability $p$ the favorite wins and with
probability $q=1-p$, the underdog wins.  We present simulation results
for $q=1/4$.

\begin{figure}[h]
\includegraphics*[width=0.425\textwidth]{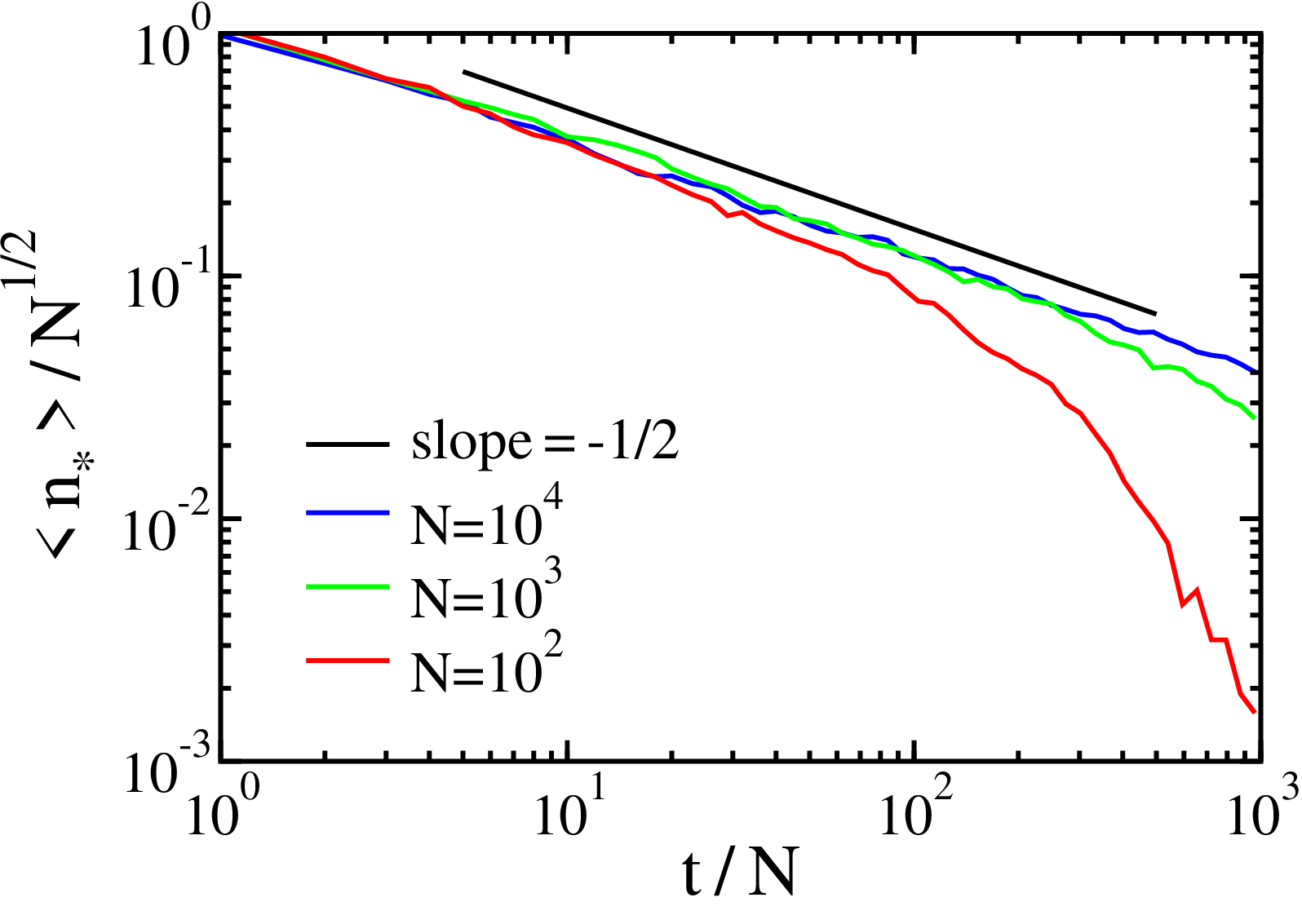}
\caption{The average rank of the champion, $\langle n_*\rangle$, of a
  league with $N$ teams after $t$ games. The simulation results
  represent and average over $10^3$ independent realizations with
  $N=10^2$, $10^3$, and $10^4$. A line of slope $-1/2$, predicted by
  Eq.~(\ref{nt}), is plotted as a reference.}
\end{figure}

The most important theoretical prediction is the relation (\ref{nt})
between the rank of the winner, the number of games, and the size of
the league. To test this prediction, we measured the average rank of
the winner as a function of the number of games $t$, for leagues of
various sizes. In the simulations, it is convenient to shift the rank
by one: the teams are ranked from $n=0$ (the best team) to $n=N-1$
(the worst team). With this definition, the average rank decreases
indefinitely with $t$. The simulations show that $n_*/N^{1/2}\sim
(t/N)^{-1/2}$, thereby confirming the the theoretical prediction
(figure 1).

\begin{figure}[h]
\includegraphics*[width=0.4\textwidth]{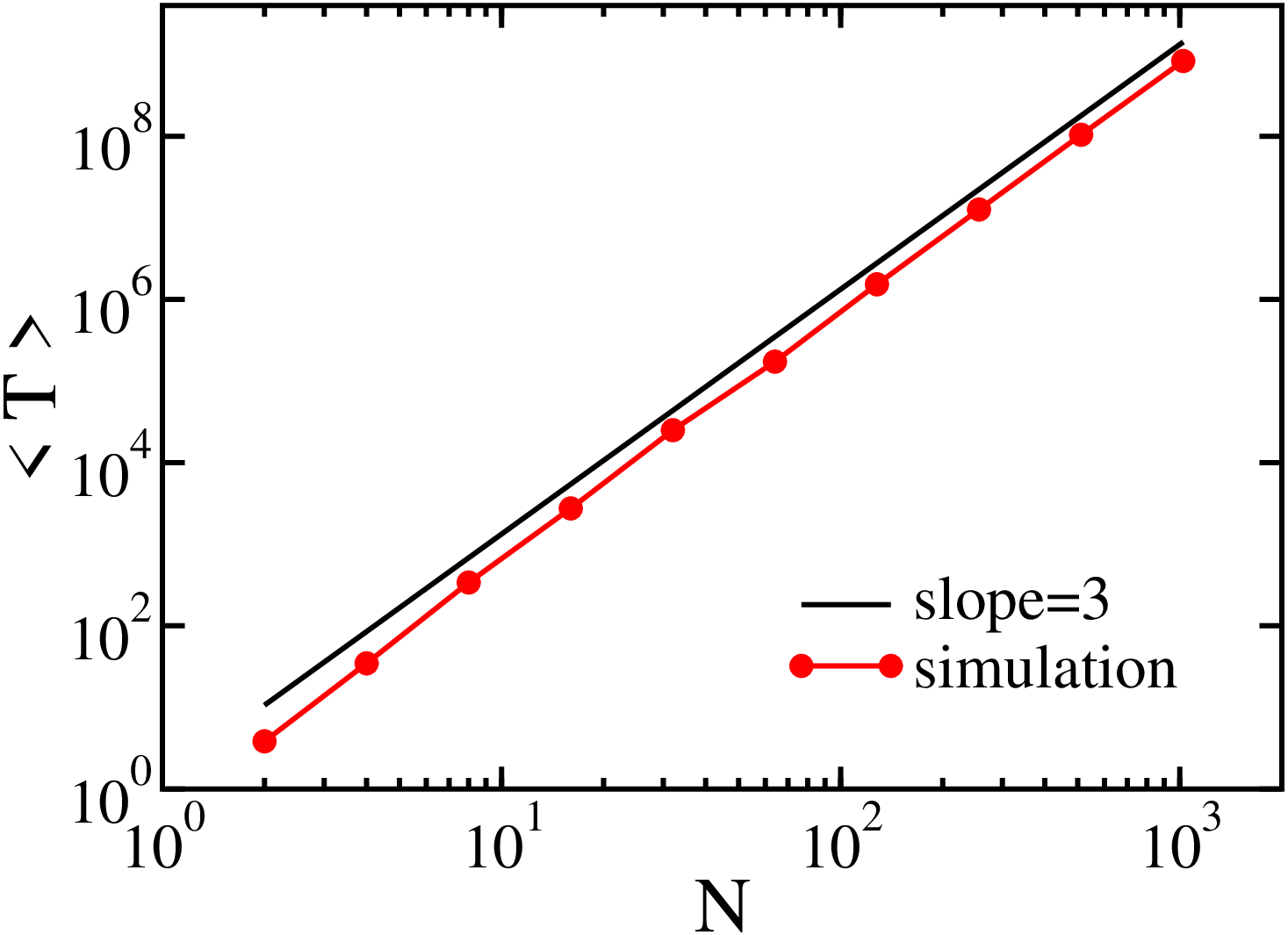}
\caption{The average number of games $\langle T\rangle$ needed for the
best team to emerge as the champion of a league with $N$ teams. The
simulation results, representing an average over $10^3$ independent
realizations, are compared with the theoretical prediction
(\ref{tn}).}
\end{figure}

To validate (\ref{tn}), we simulated leagues with a large enough
number of games, so that the best team wins with certainty. For every
realization there is a number of games $T$ after which the champion
takes the lead for good. The average of this random variable, $\langle
T\rangle$, measured from the simulations, is in excellent agreement
with the theoretical prediction (figure 2).

The simulations also confirm that the scale $n_*$ characterizes the
entire distribution as in (\ref{qn-scaling}). Numerically, we find
that the tail of the scaling function is super-exponential,
$\psi(z)\sim \exp(-z^\mu)$ with $\mu > 1$.  The observed tail behavior
is consistent with $\mu=2$, although the numerical evidence is not
conclusive.

\begin{figure}[h]
\includegraphics*[width=0.4\textwidth]{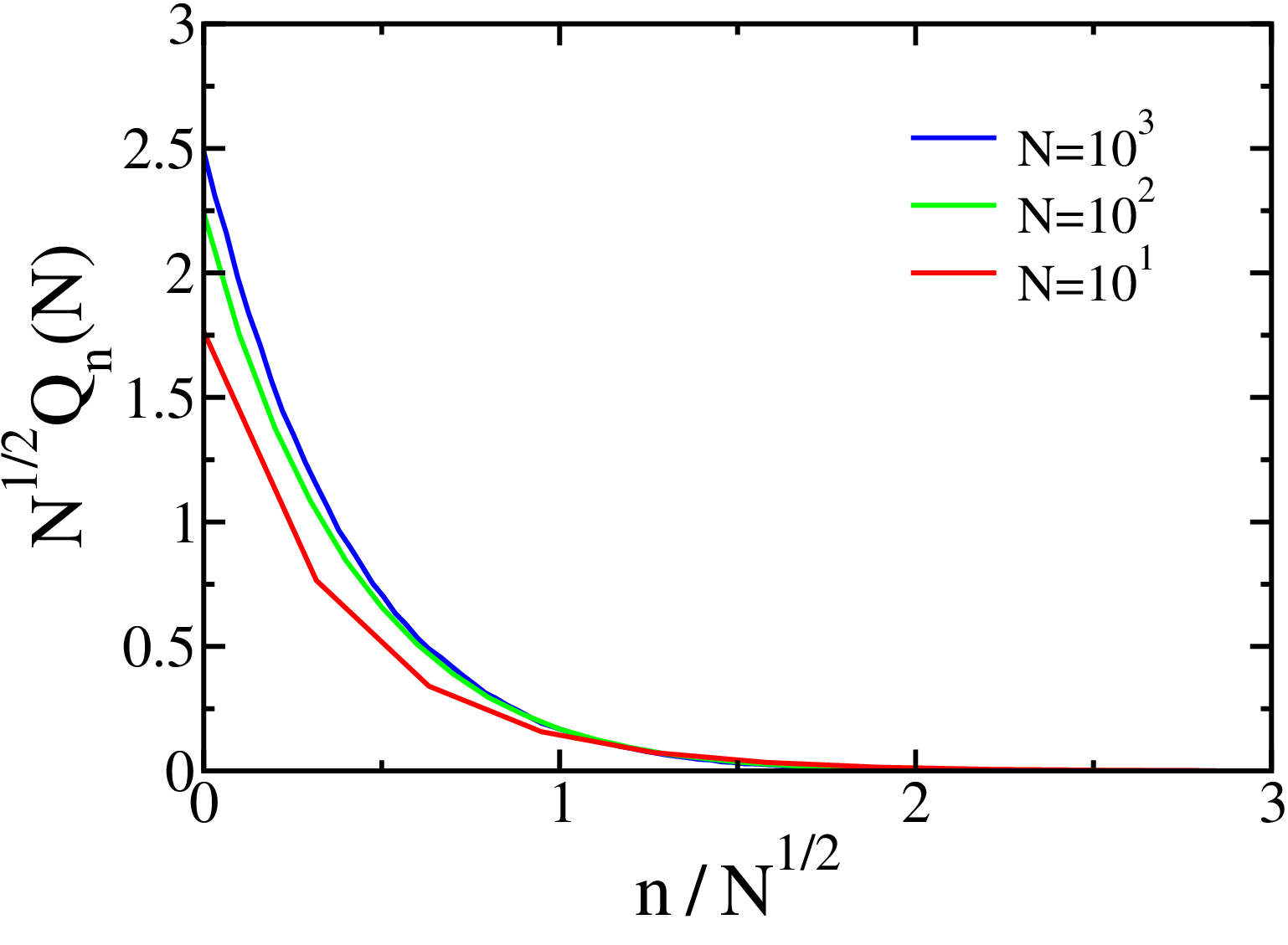}
\caption{The rank distribution of the league winner for ordinary league
format ($t=N$). Shown is the
  scaled distribution $\sqrt{N}\,Q_n(t=N)$ versus the scaling variable
  $n/\sqrt{N}$. The simulation data were obtained using $10^6$
  independent Monte Carlo runs.}
\end{figure}

To verify our prediction that multiple elimination rounds, following
the format suggested above, reduce the number of games, we simulated a
single elimination round ($k=1$). In the first stage, a total of
$N^{9/5}$ games are played.  All teams are then ranked according to
the number of wins and the top $M=N^{3/5}$ teams proceed to the
championship round. This final round has an ordinary league format
with a total of $M^3$ games.  We simulated three leagues of respective
sizes $N=10^1$, $N=10^2$, and $N=10^3$, and observed that the best
team wins with a frequency of $70\%$. The champion is among the top
three teams in $98\%$ of the cases (these percentages are independent
of $N$). As a reference, in an ordinary league with a total of $N^3$
games, the best team also wins with a likelihood of
$70\%$. Remarkably, even for as little as $N=10$ teams, the one
preliminary round format reduces the number of games by a factor
$>10$.  We conclude that the scaling results are useful at moderate
league size $N$.

\section{Imperfect champions}

Let us relax the condition that the best team must win and implement a
less rigorous championship round. Given a total of $T \sim M^c$ games
with $1 \leq c\leq 3$, each team plays $t\sim M^{c-1}$ games. From
(\ref{nt}), the typical rank of the winner scales as
\begin{equation}
\label{abc}
n_*\sim M^{\frac{3-c}{2}}.
\end{equation}

Suppose that there are infinitely many preliminary rounds.  The
analysis in Section III reveals that the total number of games scales
linearly, $T\sim M^c\sim N$, and consequently, $M\sim
N^{1/c}$. Therefore, there is a scaling relation between the rank of
the winner and the number of teams $n_*\sim N^{\frac{3-c}{2c}}$.
Indeed, the value $c=3$ produces the best champion. The common league
format ($c=2$) leads to $n_*\sim N^{1/4}$, an improvement over the
ordinary $N^{1/2}$ behavior.

If there is one preliminary round, Eq.~(\ref{t1-eq}) becomes $T_1\sim
N^{3-2\alpha_1}+N^{c\alpha_1}$ and therefore, $\alpha_1=3/(2+c)$.
Generally for $k$ preliminary rounds, the exponent $\alpha_k$
satisfies the recursion relation (\ref{alpha-eq}), and the scaling
relations $\gamma_k=3-2\alpha_k$ and $\beta_k=2(1-\alpha_k)$ remain
valid. We quote the value
\begin{equation}
\gamma_k=\frac{1}{1-\frac{c-1}{c}\left(\frac{2}{3}\right)^k}
\end{equation}
that characterizes the total number of games, $T\sim
N^{\gamma_k}$. From $T\sim M^c\sim N^{\gamma_k}$, we conclude $M\sim
N^{\gamma_k/c}$. Substituting this relation into (\ref{abc}) yields
\begin{equation}     
\label{sub}
n_*\sim N^{\nu_k},\qquad \nu_k=\frac{\gamma_k(3-c)}{2c}.
\end{equation}
Using ordinary league play ($c=2$) and one preliminary round,
$N^{3/2}$ games are sufficient produce an imperfect champion of
typical rank $n_*\sim N^{3/8}$.  Finally, we note that if each team
plays a finite number of games ($c=1$), all of the teams have a
comparable chance of winning because $\nu_k = \gamma_k \equiv 1$.

\section{Conclusions}

In summary, we studied dynamics of league competition with fixed team
strength and a finite upset probability. We demonstrated that ordinary
league play where all teams play an equal number of games requires a
very large number of games for the best team to win with certainty. We
also showed that a series of preliminary rounds with a small but
sufficient number games to successively eliminate the weakest teams is
a fair and efficient way to identify the champion. We obtained scaling
laws for the number of advancing teams and the number of games in each
preliminary round. Interestingly, it is possible to determine the best
team by having teams play, on average, only a finite number of games
(independent of league size). The optimal size of the final
championship round scales as the one-third power of the number of
teams.

Empirical validation of these results with real data may be possible
using sports leagues, for example.  The challenge is that the inherent
strength of each team is not known. In professional sports, a team's
budget can serve as a proxy for its strength.  With this definition,
the average rank of the American baseball world series champion, over
the past 30 years, equals 6. There are however huge fluctuations:
while the top team won 7 times, a team ranked as low as 26 (2003
Florida Marlins) also won.

With wide ranging applications, including for example evolution
\cite{jk,smd}, leadership statistics is a challenging extreme
statistics problem because the record of one team constrains the
records of all other teams. Our scaling approach, based on the record
a fixed team, ignores such correlations. While these correlations do
not affect the scaling laws, they do affect the distribution of
outcomes such as the distribution of the rank of the winner, and the
distribution of the number of games needed for the best team to take
the lead for good.  Other interesting questions include the expected
number of distinct leaders, and the number of lead changes as a
function of league size \cite{kr,bk}.

\noindent{\bf Acknowledgments.} We thank David Roberts for useful
discussions.  We acknowledge financial support from DOE grant
DE-AC52-06NA25396.


\begin{thebibliography}{99}
\bibitem{clean} 
  For simplicity, the team records do not carry to the 
next round.
\bibitem{small} The optimal number of rounds is finite, for a given
 number of teams.
\end{thebibliography}
\end{document}